\documentclass[twocolumn]{aastex631}

\newcommand{\citeg}[1]{\citep[e.g.,][]{#1}}

\usepackage{amsmath}
\usepackage{wrapfig}

\begin{document}

\title{Evidence for Sympathetic Flaring in TESS Data}

\author[0009-0001-8728-6894]{Veronica Pratt}
\correspondingauthor{veronica.pratt@tufts.edu}
\affiliation{Department of Physics \& Astronomy, Tufts Astronomy, 574 Boston Avenue, Medford, MA, USA}

\author[0009-0002-9757-0351]{Jason R. Reeves}
\affiliation{Department of Physics \& Astronomy, Tufts Astronomy, 574 Boston Avenue, Medford, MA, USA}

\author[0000-0002-7595-6360]{David V. Martin}
\affiliation{Department of Physics \& Astronomy, Tufts Astronomy, 574 Boston Avenue, Medford, MA, USA}

\author[0009-0005-6169-6413]{Andy B. Zhang}
\affiliation{Department of Physics \& Astronomy, Tufts Astronomy, 574 Boston Avenue, Medford, MA, USA}

\author[0009-0008-0335-3414]{Andrew Korkus}
\affiliation{Department of Physics \& Astronomy, Tufts Astronomy, 574 Boston Avenue, Medford, MA, USA}

\author[0009-0004-2529-1550]{S. Edelman}
\affiliation{Department of Physics \& Astronomy, Tufts Astronomy, 574 Boston Avenue, Medford, MA, USA}

\begin{abstract}

Most flares on the Sun occur at random, but there is a small percentage of ``sympathetic flaring'' -- the triggering of one flare by another. Previously there had been no widespread confirmation of sympathetic flares on other stars. In this work, we developed a new flare detection algorithm that is sensitive to closely-separated and overlapping stellar flares. We applied it to TESS data and discovered $\sim$ 220,000 flares on $\sim$ 16,000 stars, the majority of which are M-dwarfs. The wait time distribution between flares demonstrates an excess of closely-separated flares, relative to expectations from a Poisson process. We attribute this to sympathetic flares, occurring at a rate of between 4\% and 9\%, which matches the rate seen on the Sun. Our result is the first statistically robust detection of sympathetic flares on other stars, demonstrating a commonality between the Sun and low-mass stars.
\end{abstract}

\noindent

\section{Introduction}
Sympathetic flaring is the phenomenon in which a flare in one active region on a star's surface triggers a second flare shortly after in a nearby, but different, active region. Multiple physical processes have been proposed for the triggering of a sympathetic flare, including shockwave propagation \citep{Moreton1960,Uchida1968,Warmuth2004} and coupled magnetic field interactions between different active regions \citep{FritzovaSvestkova1976,2022Mawad}.

Sympathetic flare studies have focused primarily on the Sun \citeg{Wheatland2000, 2002WheatlandLitvinenko, 2014Tellonietal}. A powerful diagnostic tool is the wait time distribution (WTD), which is the distribution of the time in between consecutive flares. If flares are purely random (a Poisson process) then the WTD should be an exponential distribution. If the observed WTD has an excess of short wait time events relative to an exponential distribution, then this can be evidence for sympathetic flaring \citep{Moon2002,2014Tellonietal,2018Li}. There have also been observed correlations between flare wait times and the angular separation and velocity of flares \citep{2025Guite}. On the Sun, there is observational and statistical evidence for sympathetic flaring, at a rate of approximately  5\%  \citep{Moon2002,2025Guite}, with the sympathetic flares occurring on the order of one hour after the original flare.



Large photometric surveys such as Kepler and TESS have revolutionized the study of flares on other stars \citeg{2024Wainer, 2024Paudel, 2024Davenport}. For the specific study of sympathetic flaring, a key challenge is the complete lack of spatial data from the stellar-surface. Otherwise put, we know when a flare occurs but not where. In that regard, some studies discuss consider sympathetic flares as those that occur within some short time interval \citeg{Li2023,Hakamata2025}. However, we can never know if individual pair of flares is truly sympathetic. \citet{2018Li} analyzed the WTD of over 500 flares discovered in Kepler data for a single flare star KIC 11551430. The data were not well-matched by an exponential WTD.


In our paper we search for sympathetic flaring using the WTD approach. The two biggest advancements in our work are the size and completeness of our flare search. We discover $\sim 220,000$ flares on $\sim 16,000$ stars across three independently-collected samples of stars. Furthermore, the discoveries are made using a novel open-source flare detection algorithm, \textsc{toffee} (TESS Overlapping Flare Finder and Energy Evaluator), currently available for public use. \textsc{toffee} is sensitive to closely-separated flares, which many other algorithms in the literature fail to detect. \footnote{\href{https://github.com/JasonReeves702/TOFFEE}{https://github.com/JasonReeves702/TOFFEE}}

In Sect.~\ref{sec:stellar_samples}, we describe our sample of stars. In Sect.~\ref{sec: flare finding}, we introduce \textsc{toffee} and show how we discover our flares. In Sect.~\ref{sec:stats}, we discuss the statistical concepts underpinning our analysis. Finally, in Sect.~\ref{sec: results}, we present the results, before concluding in Sect.~\ref{sec:conc}.








\section{Stellar Samples}\label{sec:stellar_samples}

To search for sympathetic flares, we use three previously compiled samples of flaring stars as observed by TESS: \cite{2024Feinstein} (named Feinstein for this paper), \cite{Seli2025} (named Seli for this paper), and \cite{2025Yudovich} (named Yudovich for this paper).  In each study the authors searched for flares on 120-second cadence TESS data using a unique algorithm. Feinstein and Seli based their samples -- 3160 and 14,408 stars, respectively -- on the entire TESS catalog of $\approx 200,000$  stars with 120-second data. Yudovich used a smaller sample of 234 stars, based on an earlier study by \cite{2022Crowley}. When accounting for the overlap between the samples, our study covers a total of 16,305 unique stars and 62,450 unique lightcurves.

\begin{figure}[h]
    
    \centering
    \includegraphics[width=1 \linewidth]{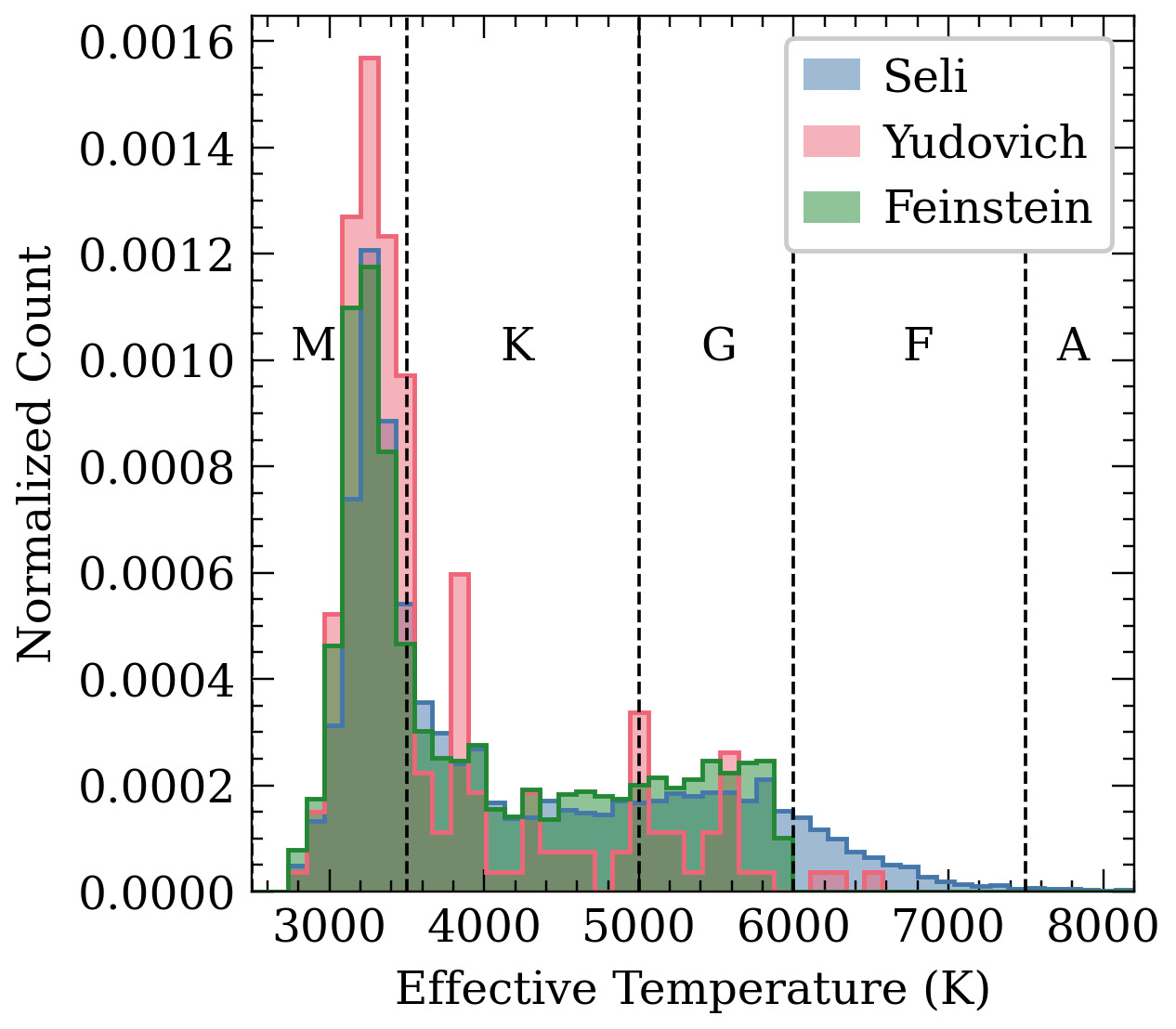}
    \caption{Normalized distribution of the temperatures of the stars across our three samples. The temperatures were taken from the TESS Input Catalog v8.2 \cite{Paegert2022}. The ranges of each stellar subtype are shown via the dashed black lines.}

    \label{fig: temp}
\end{figure}

The distributions of the kind of stars, ranging from M dwarfs to A stars, can be seen in Fig.~\ref{fig: temp}. The Feinstein sample has a sharp upper limit at 6000 K. For all three samples, M-dwarfs are the most common. This is expected, since these stars are the most common in the Galaxy and also have the highest flare rate.

These three samples provide both a list of stars and a list of flares. However, in our later analysis, we will only use the Feinstein flare sample for the purposes of finding sympathetic flares. The Yudovich and Seli flare samples are incomplete at close wait times, and hence would significantly underestimate sympathetic flaring (see Sect.~\ref{subsec:comparison_other_samples}).


\section{Finding flares with \textsc{toffee}} \label{sec: flare finding}

Here we present \textsc{toffee}, our open-source algorithm to detect flaring events. The software is currently pip installable and documented on GitHub. The motivation of this algorithm is to provide a complete sample of flaring events, including flares that overlap in time.

\subsection{TESS Data Sources} \label{sec:flare_detection_and_modeling}

 We use TESS 120 second cadence data because it is available for significantly more stars than 20 second cadence over a longer baseline. It also has an overall lower noise for flux points and thus noticeable flare signals leading to fewer false positives and easier modeling of flare shapes in the lightcurve. 


We institute a quality cut to ensure we discount spuriously bright points due to cosmic rays or hot pixels by only including flux points with a quality of 0. It has been shown that flux points with a quality of 512 can be the peaks of flares, with studies including points of quality 0 and 512 in order to have a complete sample of flares across all amplitudes \cite{2020Feinstein}. To maintain a pure sample of flares we elected to use the more conservative cut of only selecting points with a quality value of 0.

When discovering flares with \textsc{toffee} we wanted our results to be directly comparable to the flares discovered by \citet{2024Feinstein,Seli2025,2025Yudovich}. Therefore, we utilize only the same TESS sector coverage as used in each paper. So, for the stellar sample adopted from Feinstein, we utilize TESS Sectors 1 through 67, for Yudovich TESS Sectors 1 through 55, and for Seli TESS Sectors 1 through 69.

\subsection{Lightcurve Detrending} \label{sec:detrending}

\begin{figure}
    \centering
    \includegraphics[width=\linewidth]{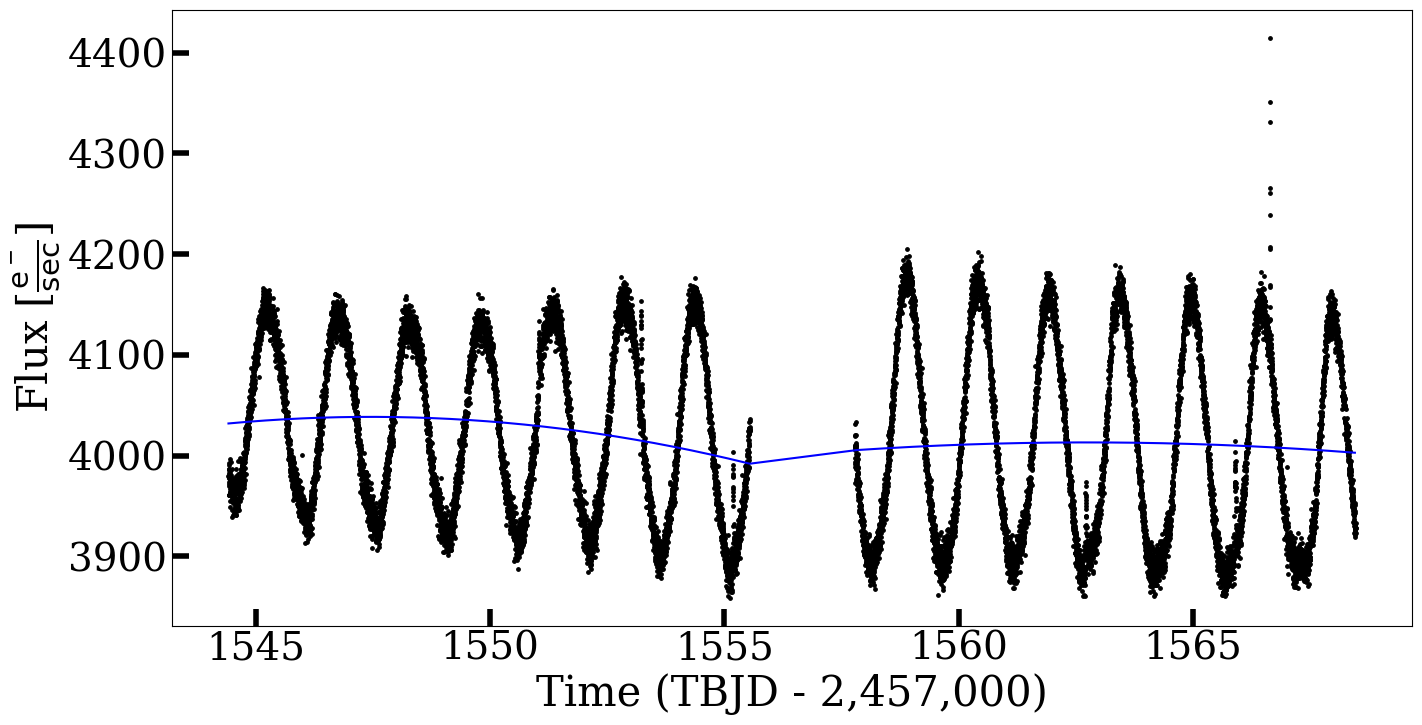}
    \includegraphics[width=\linewidth]{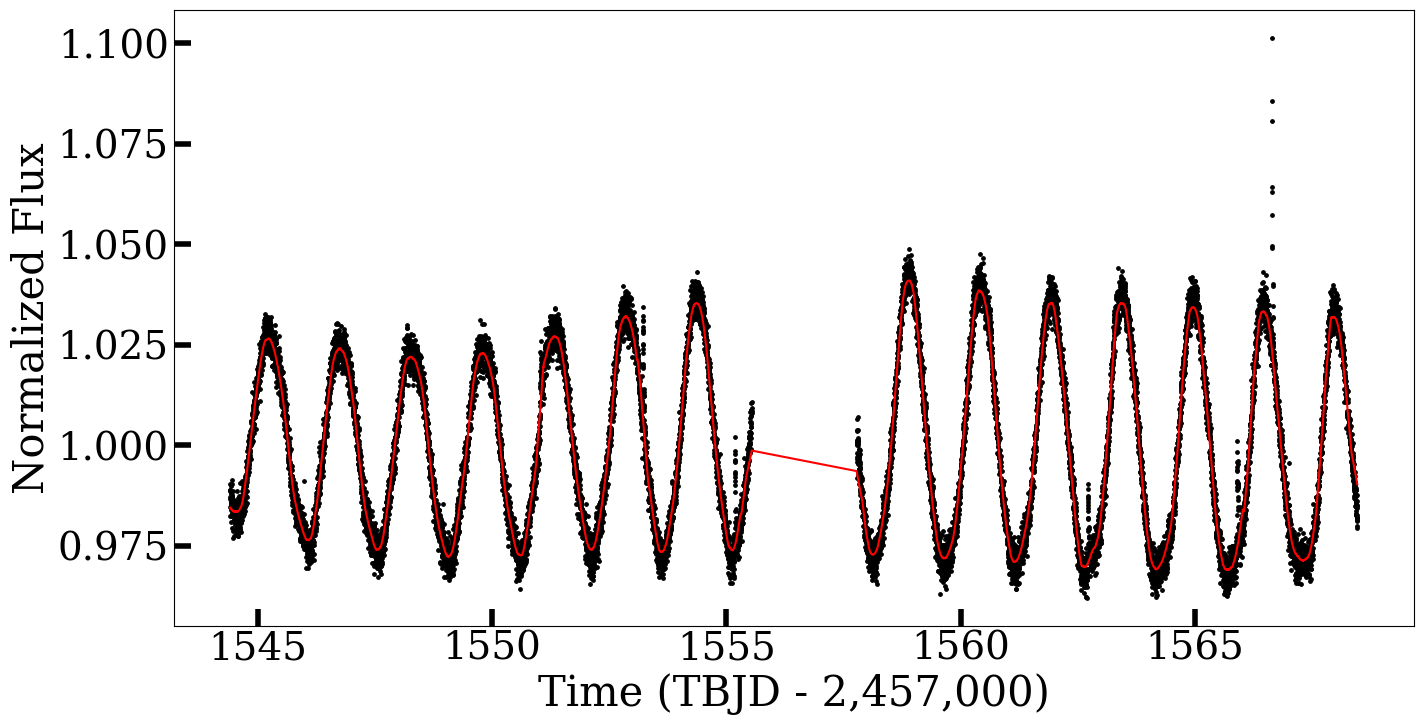}
    \includegraphics[width=\linewidth]{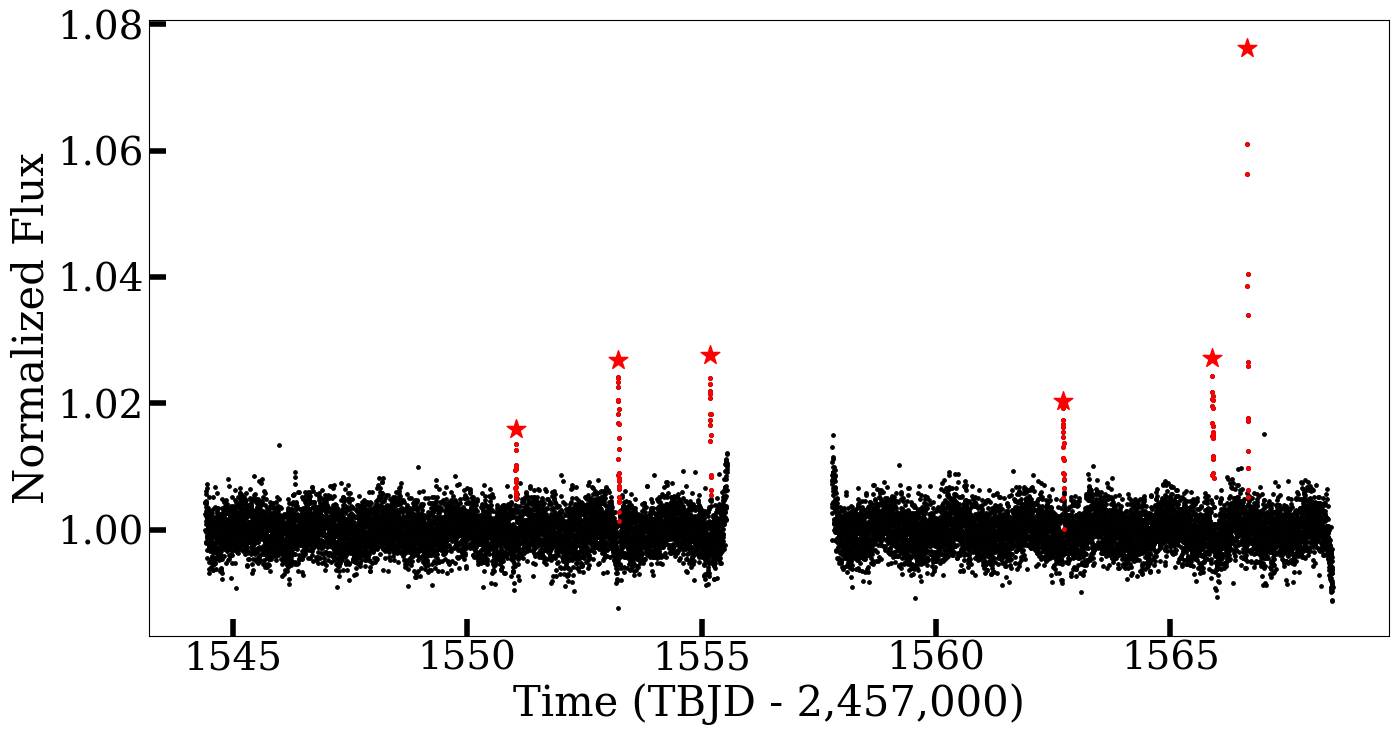}
    \caption{Detrending and flare search process from raw lightcurve (\textit{top}) to final flattened curve (\textit{bottom}). \textit{Top} shows the raw time resolved brightness of the star in black points with the quadratic trend coming from the orbit of the TESS telescope in blue. \textit{Middle} shows the quadratic subtracted lightcurve overlaid with the trend found by \texttt{wotan} in red. The lightcurve shown in \textit{bottom} is the final flattened lightcurve in black and the detected flares. Primary flares are colored in red with the peaks labeled as red stars and the secondaries labeled in blue with a blue star representing its peak. We also add an inset zoom-in around the secondary flare to show its morphology. Residual signal of the spot modulation is seen in the final lightcurve shown in \textit{bottom} which leads us to consider the global spread of the points to calculate the flux threshold as opposed to the photometric error. We also note that cutting 100 cadences on either side of a break prevents classifying detrending atrifacts as flares as seen in the orbit break of this lightcurve.
    }
    \label{fig:detrending}
\end{figure}

The aim of our flattening is to remove any non-flare instrumental and astrophysical trends, particularly the effects of spot modulation and stellar pulsation that may negatively impact the efficacy of our flare detection. We segment each light curve by TESS orbit, fitting and dividing by a quadratic trend using \texttt{numpy}'s polyfit function to remove any instrumental trends on the timescale of a TESS orbit (13.7 days). We then utilize Tukey's biweight filters \cite{1977Tukey} provided by the \texttt{Wotan} package \cite{2019Hippke} applied to our light curves with a standard $0.25$ day window length, a threshold below which we observed noticeable attenuation of flares by the flattening algorithm harming potential detection.

Next, a Lomb-Scargle periodogram is created per TESS orbit of data. If the false alarm probability (FAP) was above $20\%$, we deem any periodicity in the light curve to be sufficiently flattened. Otherwise, we conduct further flattening with \texttt{Wotan}'s `median' filter on a window length equal to half the periodogram's peak signal period. This procedure is then repeated iteratively on the new light curve until either the FAP has reached above $20\%$, or the window length reaches 0.1 days. We choose not to detrend more aggressively than this since we do not want to attenuate the depths of flares. Lastly, we apply a mask to each lightcurve after detrending to cut off 100 points on either side of each break. This is to account for instrumental effects that frequently occur near gaps in TESS data. Fig. \ref{fig:detrending} demonstrates the process from taking a raw lightcurve, through the bi-weight flattening, before labeling the detected flares using \textsc{toffee}.

\subsection{Primary Flare Detection} \label{sec:flare_detection}

\begin{figure*}
\centering
\includegraphics[width = 0.49\linewidth]{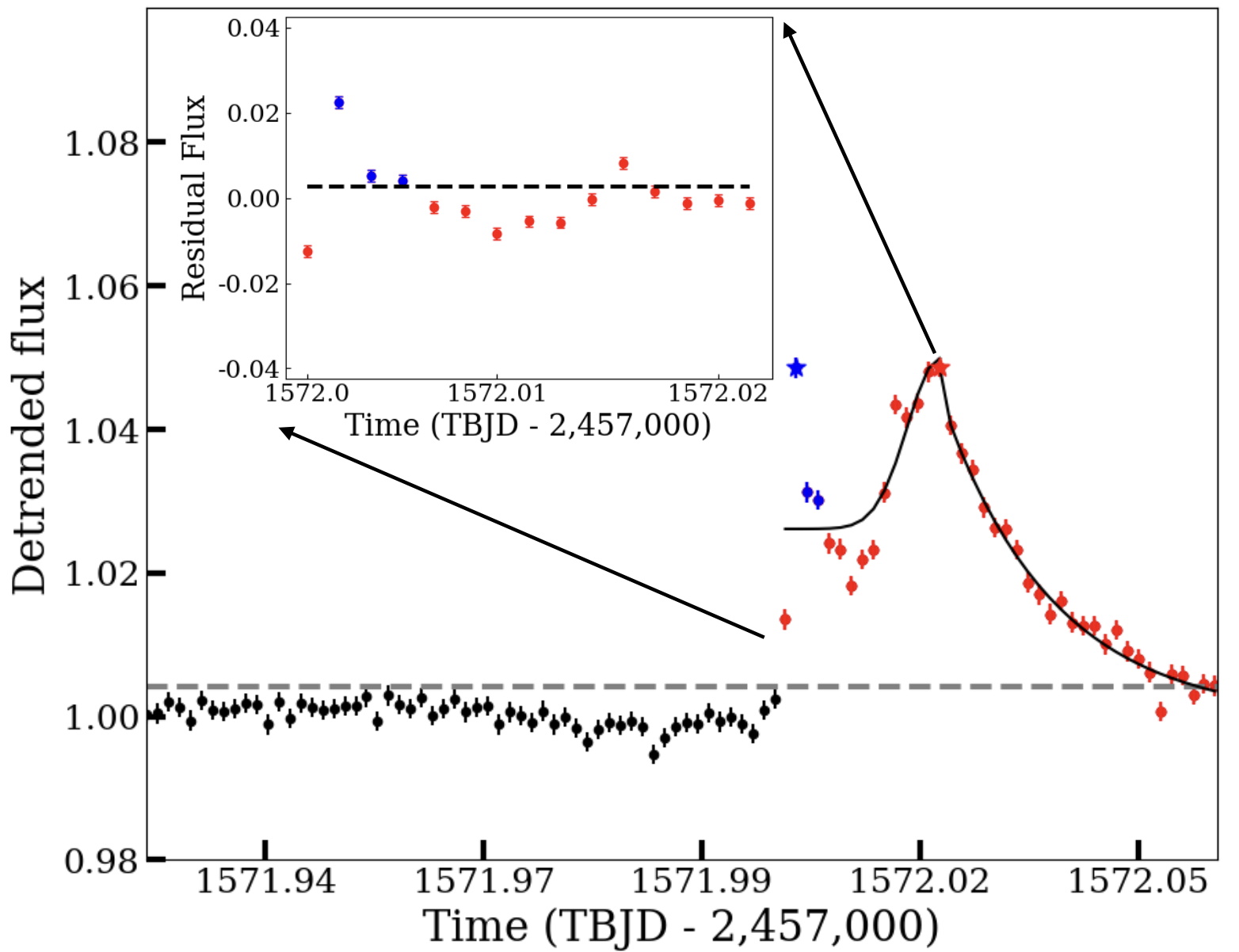}
\includegraphics[width =  0.49\linewidth]{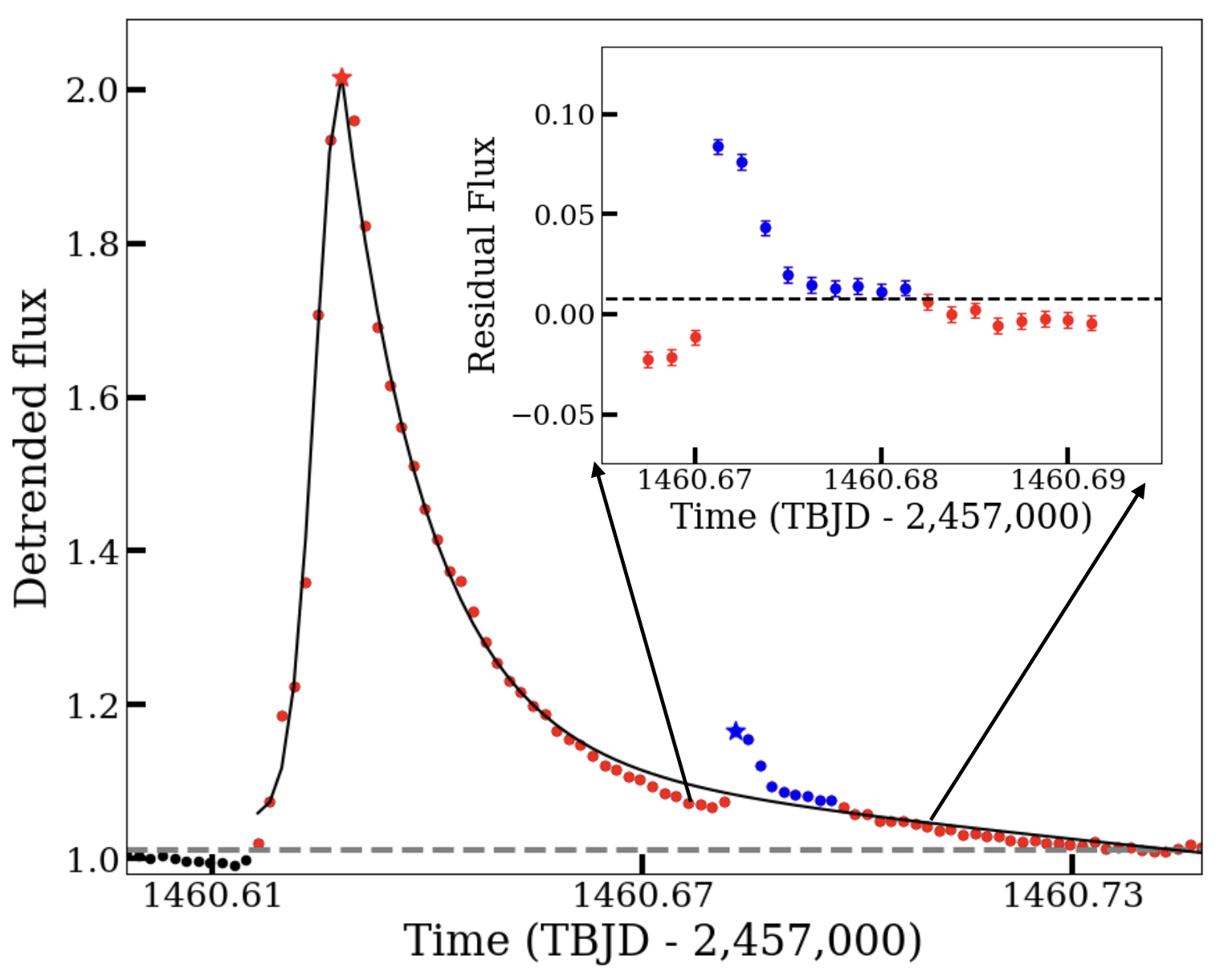}
\caption{Demonstration of flare detection and modeling of \textsc{toffee} used to find secondary flaring events. \textit{Left} demonstrates an example of a secondary flare event being found at an earlier time compared to the primary coming from TIC 295777692 Sector 10, and \textit{right} demonstrates an example of a secondary flare being found at a later time compared to the secondary coming from TIC 100481123 Sector 5. Non-flare flux points are colored in black with a gray dashed horizontal line representing the $3\sigma$ threshold of the global spread used to find flares. The large primary flares with points identified as being associated with the flare are colored in red and points associated with the secondary are colored in blue with the peaks labeled with stars. The black line traces the best fits of the Gaussian rise and the double exponential decay models used to model the rise and the decay of the flare, respectively. The inset figures on each figure show residuals which are used by \textsc{toffee} to identify a secondary flare by  finding the points above the $2\sigma$ threshold shown in a black dashed line. \textit{Left} shows the residuals to the primary flare model in the rise portion of its respective flare. We note the Gaussian rise fitting procedure does not require the function to pass through the first point in the flare in order for secondary flare detections to be robust. \textit{Right} shows a portion of the residuals in the decay portion of a flare around the detection of a secondary flare.}\label{fig: toffee}
\end{figure*}

\textsc{toffee} is a threshold-based method like \cite{2021Ilin,2014Davenport,2019Yang,2022Howard}. The code finds all the points above a flux threshold given by the spread in the lightcurve, and then works down from the brightest points to the least bright above the threshold, modeling around each point to determine to which specific flaring events the bright points potentially belong. 
The median of the post-flattened lightcurve is subtracted from the flux to center around zero, then the 84th percentile rank value is taken to represent one standard deviation above the typical quiescent flux, which is subsequently multiplied by three to calculate the $3\sigma$ threshold for detection. We use the spread of the flattened lightcurves rather than the photometric error of each flux point to account for possible residual spot modulation, inflating the errors to require the fluxes to be slightly larger in order to be counted as a flare.

All flux points above a $3\sigma$ threshold are labeled as flare candidates. The code then sorts the points above $3\sigma$ in descending order of flux and iteratively examines each to see if it meets the following criteria of a flare. First, the code checks to ensure there are a sufficient number of bright points above the $3\sigma$ limit to discount the possibility of the reading being randomly bright. Around each event the code looks to the left at earlier times and to the right at later times to determine where the flaring event begins and ends, and it determines which of the other bright points belong to the potential flaring event. A flare event is determined to have ended when three consecutive points lie less than $2\sigma$ above the median flux.
Points below the $3\sigma$ threshold can still be considered a part of the flare event, so long as they lie between two points that are above the $3\sigma$ threshold. This represents potential areas of flare decay that fluctuate about the threshold due to Poisson noise but are still truly a part of a flare.

In order to better capture low amplitude flares, we relax the requirement used in \cite{2021Ilin} \& \cite{2015Chang} where bright points need to be consecutive. Instead, we require the possible flaring event have three in four flux measurements above the $3\sigma$ limit. The concern of allowing few cadences to determine flare detection is possible contamination from cosmic rays, however ultimately requiring three points above the flux threshold makes a robust case for any given signal being a flare. Such a requirement also is seen to be effective in eliminating false-positive detrending artifacts where epochs of the lightcurve rise and fluctuate around the $3\sigma$ threshold but fail to have 75\% of such points above the flux threshold. To counter the potential side-effect where a true flare occurs on the rise of residual red noise such that the epoch of the lightcurve surrounding the flare fluctuates around the threshold we add an additional check that if there are three consecutive points above $4.5\sigma$ the epoch is still considered a valid flare detection. 

\subsection{Secondary Flare Detection}

What distinguishes \textsc{toffee} from other threshold-based algorithms is the ability to find secondary flares that overlap in time with the primary flare. This feature makes us complete at short flare wait times and hence makes us sensitive to potentially sympathetic flares. In Fig. ~\ref{fig: toffee}, we demonstrate an example of a \textsc{toffee} detection of a secondary flare overlapping with the primary.

After finding the initial ``primary'' flare, the code fits a flare model of a Gaussian rise function \cite{2014Pitkin} and a double exponential decay \cite{2014Davenport}. For both the rise and the decay we use \texttt{curve\_fit} from the Python package \texttt{scipy}, utilizing the time and flux coordinates of the lightcurve and the $1\sigma$ global spread of the photometric data as the error. For the rise, the Gaussian model is fit with the requirement that it must pass through the peak of the primary flare. We calculate the residuals of the fit as the difference between the fitted values from the values of the measured flux values at the same times. If three consecutive points with residuals above $2\sigma$ of the global spread are found, then the code concludes there is an additional, secondary flare, to the left of the primary: we call this the secondary because it is found second in the code due to its lower flux. Due to the usage of 120 second cadence data, features of a small amplitude flare occurring during the rise of a large flare rarely shows up as a detectable signal in the residuals. However, such a search at lower times of the peak of a bright flare helps to determine if there is another flare that started before the later, larger flare and happens to overlap in time. Such an example is shown in Figure \ref{fig: toffee}. The lower $2\sigma$ limit has been previously found to be appropriate for flare detection in lightcurve data \citep{2023Rivera}. To account for the background flux of the primary flare the amplitude of the secondary as the difference between the flux of the secondary at its peak and the flux of the fitted model.

The decay portion works in a similar manner. A double exponential is fit to the decay of the primary. If three consecutive points are found to have residuals above the $2\sigma$ threshold limit, we claim it as a secondary detection. It is possible for multiple distinct epochs to exist in the residuals indicating additional ``tertiary" or higher order flares. \textsc{toffee} contains functionality to find any number of secondary flares in the decay portions of flaring events. However, such functionality has been turned off for the purposes of this study in the interest of maintaining a simple sample of secondary flares. Incorporating high order secondary flares into the analysis was seen to increase the measured rate of sympathy, but we opted to use a more conservative method of detecting overlapping flare events.

Through testing and visual inspection, we discovered that misclassifications sometimes occurred in flares with a high signal-to-noise ratio, such that the double exponential decay curve does not sufficiently capture the rapid decay portion of the lightcurve immediately after the peak. This caused the residuals to be high enough to trigger a secondary flare ``discovery'', despite there being no obvious secondary flare visually. To avoid this potential false positive, we add a requirement that the secondary flares of the decay begin at least ten minutes (five cadences) after the primary. A similar problem arose with very bright flares that have a normalized flux amplitude greater than 2. In these events inner complexities showed up as large, significant deviations from the double exponential decay that the code flagged as secondary flares \citep{2022Ward}. 
To ensure a secondary is true and not the result of inner complexities of the primary, we stipulate that for any flare with an amplitude of 2 or greater, the threshold requirement for a secondary  to be a flare on the rise or decay is raised to $10\sigma$. This eliminated false positives but allows for secondaries provided they are sufficiently bright to not be largely obscured by the primary. We also note that typical flare frequency distributions result in many more low amplitude flares than high amplitude flares, and hence a normalized amplitude of 2 or greater is a rare occurrence.

\begin{figure}
    \centering
    \includegraphics[width=\linewidth]{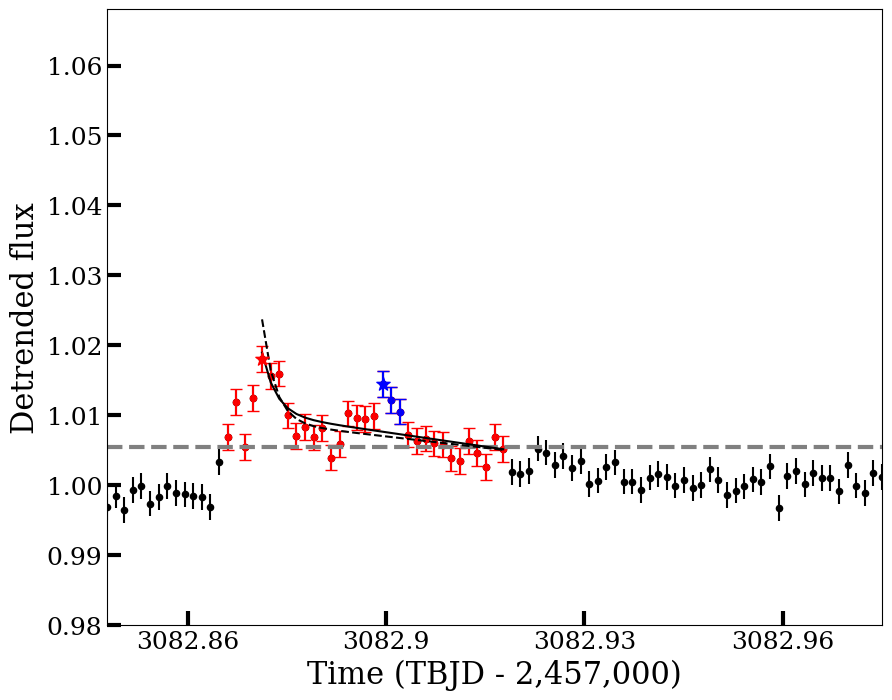}
    \caption{Example of refitting decay portion of primary flare to find secondary. The \textbf{solid} line shows the first attempted fit using all flux points in the decay. The three blue points represent the flux points labeled as being marginally missed as a flare with residuals $\geq 1.5\sigma$. The \textbf{dashed} line shows the refitted double exponential decay after removing the blue points lessening their effect on the fitting. After refitting the new residuals of the three points lie above the $2\sigma$ threshold for recognition as a secondary and is thus added to the flare catalog. This shows the general trend that secondary flares recovered via a second fitting are generally low amplitude flares very close to $2\sigma$ signals.}
    \label{fig:refitting_secondaries}
\end{figure}

In modeling the decay of the primary flare the quality of the fit will be affected by the presence of a secondary flare. This makes finding secondaries of low amplitude much harder, as their residuals are lowered via their affect on the result of \texttt{curve\_fit}. To account for this \textsc{toffee} finds potentially missed secondaries defined as a series of at least three consecutive points in the decay with residuals above the $1.5\sigma$ deviation from the global spread in the data. The algorithm then clips out those points that potentially affected the fit and redoes the \texttt{curve\_fit} regression before adding them back in. If there are now three consecutive above the two-sigma limit they are added as a new secondary flare. An example of a secondary flare added via this feature is shown in Fig. \ref{fig:refitting_secondaries}. This feature adds 55 additional flares from Feinstein, 33 additional flares from Yudovich, and 408 additional flares from Seli in this study. 

\subsection{Measuring Flare Properties}

\textsc{toffee} records the flare start and end times, flare peak time, amplitude, equivalent duration, and a series of flags on how the flare was detected (see Sect. \ref{sec:flare_flags}). The flare start and end times are found as described in the previous section. The peak times are found as the points in the flare with the highest flux. The amplitude is found as the highest flux point in the flare minus the median flux of the star.

For equivalent duration, defined as the amount of time in which the star would need to shine in its quiescent power to release the energy produced by the flare, the typical method of integrating the lightcurve of a flare is utilized. For primary flares this requires subtracting the median flux during the event of a flare to isolate the emission of the flare by itself and finding the area under the curve using trapezoidal rule via \texttt{trapz} from the \texttt{numpy} package. For secondary flares, in order to account for the background emission from the primary flare, the area under the residual is used where the first point above the two-sigma limit and last point above the two sigma limit identified as a part of the flare are used as the beginning and end, respectively. The equivalent durations are recorded in units of seconds. Multiplying the equivalent duration by the luminosity of the star calculates the energy of the flare. 

\subsection{Flare Flags} \label{sec:flare_flags}

To test the effects of various attributes of the flares on the results, \textsc{toffee} keeps track of other qualities of the flare, including the total number of points associated with the flare, the number of points above the threshold, the amplitude of the flare in terms of the standard deviation of the global spread, and a flag of whether or not the flare was found as a primary, a primary where a refit was attempted but did not find a secondary flare, a secondary that was found on the first attempted regression, or a secondary that was only found after refitting the decay. 

For the Feinstein sample, 18,905 flares are labeled as a primary flare and 676 are labeled as a secondary flare for a total rate of 3.5\% of all flares being secondaries. For the Yudovich sample 14,013 primaries and 533 secondaries are found for a 3.7\% rate of secondaries. \textsc{toffee} found 187,222 primaries and 5,648 secondaries from the Seli sample for a secondary rate of 2.9\%. 



\begin{figure}
    \centering
    \includegraphics[width=\linewidth]{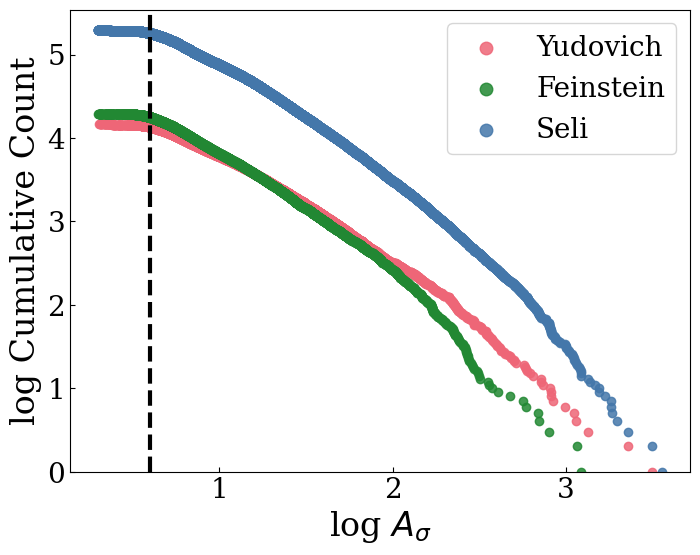}
    \caption{Cumulative amplitude counts of flares in the Feinstein (green), Yudovich (pink), and Seli (blue) samples. Flares amplitudes are expressed in terms of the spread of the lightcurve, $\sigma$, $A_{\sigma}$, or higher expressed in logspace. Deviations from a straight line in log-log space (a power-law) are emblematic of incompleteness at low amplitude (left of the vertical dashed line) or small number statistics at high amplitude.}
    \label{fig:flare_completeness}
\end{figure}

Worthy of consideration for this study is incompleteness not only as a results of missing low amplitude flares with amplitudes $< 3\sigma$, but also the incompleteness due to residual spot modulation. If flattening cannot eliminate all variability, then the catalog will be incomplete for small amplitude flares that occur during the trough in the spot modulation flux. This incompleteness would bias the wait time distribution to make flares more likely discovered in clumps near the peak of the spot modulation. To examine the role of incompleteness, we visualize the cumulative number counts of the flares based on their amplitudes in terms of $\sigma$, shown in Fig. \ref{fig:flare_completeness}. For all three samples considered the trend deviates from the expected straight line in log-space at $\log{A_{\sigma}} \approx 0.6$ which corresponds to an amplitude of $A_{\sigma} \approx 4$. This cut is consistent with the notion that in lightcurves with residual red noise $1\sigma$ will trace the amplitude of the modulation, thus adding it to the original threshold of $3\sigma$ will create a completeness criterion. The additional $4\sigma$ cut is applied only to primary flares because the detection of secondaries is independent of the bias resulting from residual spot flux modulation. As an assurance check we apply the $4\sigma$ cut to all flares, primary and secondary, and refit the Weibul distribution. We note a rise in the best-fit k-values for the samples, however the results still remain significantly below $k=1$.

\section{Statistical Methodology} \label{sec:stats}

\subsection{Wait Time Distributions}\label{subsec:WTDs}

The simplest model of flare occurrence is a stationary Poisson process: flares on a given star occur independently of each other at a constant rate, denoted by $\lambda$. In this case, the wait time distribution (WTD) of the time $t$ between each successive flare follows an exponential distribution:

\begin{equation}
f(t) = \lambda \mathrm{e}^{-\lambda t}.
\end{equation}
In this context, the average wait time between flares is  $<t>=1/\lambda$. A WTD drawn from a Poisson distribution is always going to be at its maximum at the shortest measurable wait time. In other words, coincidental close flares are common, even in the absence of causality.

We argue that there are three primary causes for a star's observed WTD to depart from an exponential: 

\textbf{1. Causality between flares}. The occurrence of one flare changes the likelihood of a subsequent flare. Since ``sympathetic flaring'' is the increased probability of a subsequent flare, those events would create an excess of short wait-time events. Stars may also exhibit the opposite effect; for example, the occurrence of a flare reduces the probability of a subsequent flare. This opposite phenomenon would therefore create a dearth of short wait-time events. We dub this ``reload flaring,'' i.e. the star has to ``reload'' after a flare.

\textbf{2. Non-stationary Poisson process}. The flares occur independently, but the governing process is more complex. For example, if the star's underlying flare rate changes over time, due to an activity cycle akin to the Sun's 11 year cycle, that would create a nonstationary Poisson process. This would cause the WTD to deviate from an exponential, potentially mimicking the presence of sympathetic flares. To address this potential degeneracy, we analyze the data on a year-by-year basis, since a shorter time frame means the flare rate is more likely to be constant across it. Alternatively, different parts of the star may flare at different rates, so there might be a combination of multiple stationary Poisson processes. This phenomenon alone would not mimic sympathetic flaring; the combination of two Poison processes with rate parameters $\lambda_1$ and $\lambda_2$ is also a Poisson process with rate parameter $\lambda_1+\lambda_2$. However, if different parts of the star flared at different rates, even if those rates were independently static, the rotation of the star would bring different flaring regions in and out of view, on a timescale of $\sim$ days to tens of days, inducing an observed non-stationary Poisson process, and hence could mimic sympathetic flaring. We do not believe this effect to be a concern though, due to multiple past studies finding no correlation between flare rates and rotation phase \citep{HuntWalker2012,2014Hawley,2015Lurie,2018Doyle,2019Doyle,2020Doyle}. The most recent such study was by \citet{Zhang2025}, which was conducted by largely the same team as this paper, using the same \textsc{toffee} algorithm and on the same sample of stars.

\textbf{3. Detection incompleteness}. There may be biases in the flare detection itself. In prior studies, these issues typically lead to a scarcity of short wait-time flares, since it is more difficult to detect flares that occur shortly after each other. To address this problem, we wrote the \textsc{toffee} algorithm.

To understand the intervals between successive flares in our data, we model the WTDs of our catalogs with both an exponential fit and with a slightly more complex distribution that is still in the same family as an exponential: a Weibull distribution. Weibull distributions are frequently used to model interlinked events, or events with a certain amount of memory, hence this is why they have been applied to the study of sympathetic flares \cite{2018Li,2014Tellonietal}. The probability density function (PDF) of a Weibull WTD is given by: 

\begin{equation} 
f(t) = \lambda k(\lambda t) ^{k-1} e^{(-\lambda t)^k}, \end{equation}
where $\lambda$ is the rate parameter and the new variable $k$ is the shape parameter. 

\begin{itemize}
    \item $k < 1$: Excess of short wait times, as expected for sympathetic flaring
    \item $k = 1$: Weibull distribution reduces to an exponential distribution, as expected for a stationary Poisson process
    \item $k > 1$: Dearth of short wait times, as expected for detection incompleteness
\end{itemize}

\subsection{Calculating Flare Wait Time Distributions} \label{sec: wait times}

There are several kinds of gaps in TESS data. Some gaps are when observations are not occurring, e.g. during data download and safe mode pauses. Some gaps come from the removal of low quality data, e.g. spacecraft momentum dumps, cosmic rays, and scattered light from the Earth or Moon. When calculating the flare WTDs, we take into account all of these data gaps, and we do not count any wait time in our calculations if there is a data gap longer than 14 minutes. For the calculation, we use the time of the highest point of the flare, denoted as \textit{tpeak} in our flare catalog. 



Throughout our statistical analysis, we frequently employ bootstrapping so that we can determine a distribution of various parameters. For a given star, we calculate a bootstrapped WTD as follows. First, we take the initial WTD, which was created as mentioned above. From that initial WTD, we resample the flare wait times with replacement 1,000 times, as opposed to the resampling the flare times, which would lead to a false spike of wait times of zero. We subsequently re-perform all calculations for calculating $k$ parameters and sympathetic excesses on each of those 1,000 samples.

\begin{figure*}
    
    \centering
    \includegraphics[width=1 \linewidth]{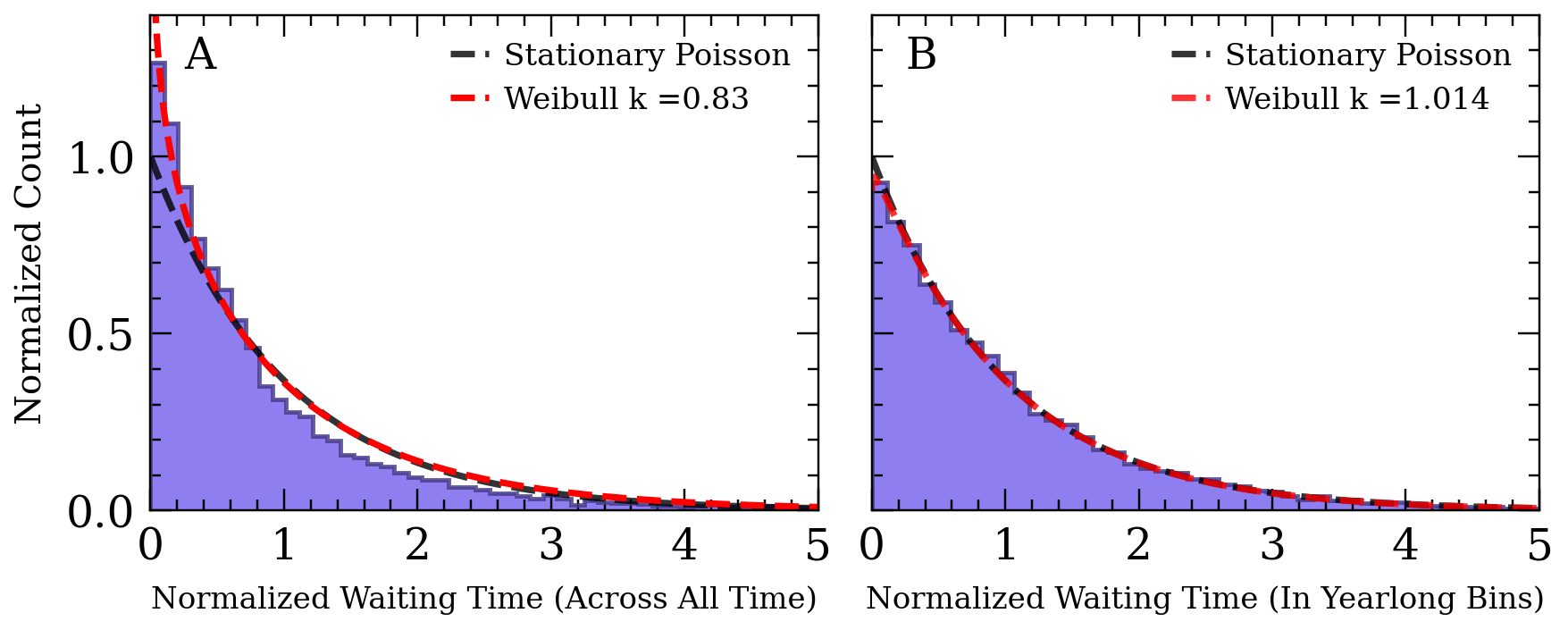}
    \caption{The WTD Probability Density Function for two different normalization schemes of the same nonstationary simulated stellar flare rates. In this simulated setup there are no sympathetic flares. A) The purple WTD is the normalized waiting times without any time bins, or across all time. The red dashed line is the best-fit Weibull distribution for the generally normalized waiting times, with $k=0.828$. The black dashed line is a stationary Poisson WTD. B) The purple WTD is the normalized waiting times, normalized in bins of one year. The red dashed line is the best-fit Weibull distribution for the yearly normalized waiting times, with a  $k= 1.012$, close to the value of 1.0 expected for purely coincidental flares. The black dashed line is a stationary Poisson WTD.  }

    \label{fig:norm_wait_times}
\end{figure*}

\begin{figure*}
\centering
\includegraphics[]
{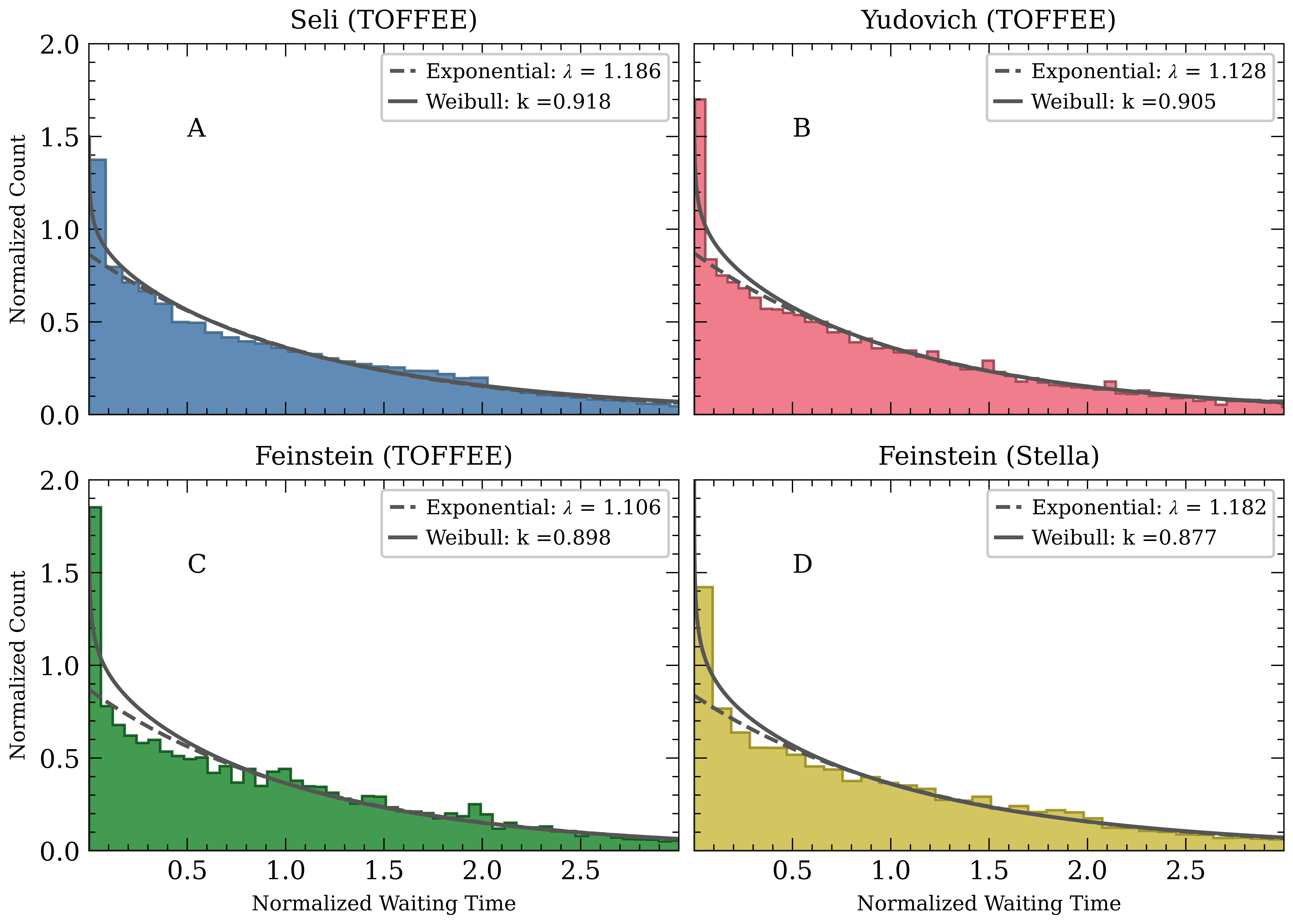}
\caption{WTDs of each catalog created by \textsc{toffee}. The WTDs are constructed from the wait times of each star by dividing by the average flare rate per star and then stacking together all data within a given sample. We accounted for potentially variable flare rates on a given star by doing the normalization on yearlong segments of data. A) The WTD of the catalog from Seli, created by \textsc{toffee}, with the dashed gray line representing the underlying exponential distribution fit to the WTD and the rate parameter highlighted in the legend. The solid gray line shows the Weibull distribution built with that rate parameter from the Poisson fit, with the best fit \textit{k} parameter highlighted in the legend.  B) The WTD of the catalog from Yudovich, created by \textsc{toffee}, with the same lines as in part (A). C) The WTD of the catalog from Feinstein, created by \textsc{toffee}, with the same lines as in part (A). D) The WTD of the flare catalog created by \texttt{stella}, as a comparison to test \textsc{toffee}'s completeness, with the same lines as in part (A). }\label{fig: WTDs}
\end{figure*}

\begin{figure*}
\centering
\includegraphics[width = \textwidth]
{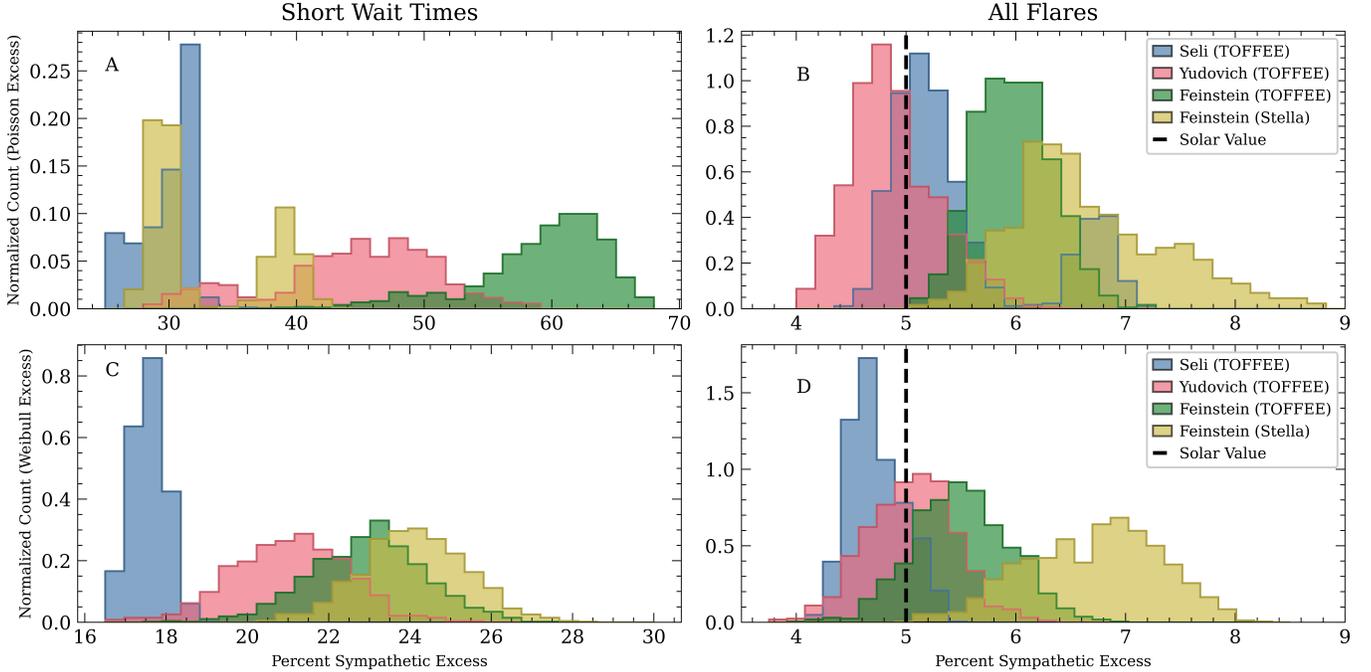}
\caption{The percent sympathetic excesses of each catalog created by \textsc{toffee}, and for comparison the \texttt{stella} catalog from Feinstein \cite{2024Feinstein}. A) The percent excess as calculated from the deviation from a Poisson fit, with the $\lambda$ parameter fit to the Poisson region of the WTD, at short wait times. Short in this case is defined as where the data begins to deviate from a Poisson distribution. B) The percent excess as calculated from the deviation from a Poisson fit across all wait times, calculated by dividing the short excess in the Poisson calculation by the total area under the histogram curve. C) The percent excess as calculated from the difference between the area under the Weibull WTD fit and the area under the Poisson WTD fit for the perviously defined short wait times. D) The percent excess as calculated from the difference between the area under the Weibull WTD fit and the area under the Poisson WTD fit for all wait times.} \label{fig: excess}
\end{figure*}

\subsection{Calculating Sympathetic Excess}\label{subsec:calc_sympathetic_excess}

Our evidence of sympathetic flaring comes from an excess of short wait time events relative to the exponential WTD expected for a stationary Poisson process. Alternatively, an excess of short wait time flares could be indicative of a variable-rate Poisson process. 

A nonstationary Poisson process could inflate the amount of short wait-time flares relative to an exponential distribution, masquerading as sympathetic flaring. To test ways to combat this issue, we inject flares (modeled from \citealt{2014Davenport}) on 50 stars over 8 years according to a Poisson process with variable flare rates. The flare rates varied according to linear and sinusoidal functions. We then normalize the data per star in two ways: 1) our typical method of yearlong normalization, as applied in the rest of our study, and 2) over the entire 8 years. The data and best fit models with a Weibull distribution are shown in Fig.~\ref{fig:norm_wait_times}.

When we fit the Weibull distribution to the WTD \textit{not} normalized in yearlong bins, the best-fit \textit{k} parameter was $k=0.828$, which could be confused as a sign of sympathetic flaring. When we instead normalize in yearly bins, the WTD is best-fit by a  Weibull with $k=1.012$, which is essentially an exponential distribution, as expected for a stationary Poisson process. This change demonstrates that the yearly binning of data allows us to account for time-dependent, non-stationary, flare rates. Yearlong binning is also robust in accounting for stars having  \textit{simultaneous} multiple Poisson processes at different rates. When we group together our modeled stars of different flare rates in pairs and repeat our procedure, we again find that with yearly normalization we have properly corresponding k parameters given the amount of sympathetic flares.

Our actual observed WTD (Fig.~\ref{fig: WTDs}) visually looks like a disjoint combination of an exponential distribution plus a very narrow sympathetic excess at short wait times. Contrastingly, a non-stationary Poisson process, like the blue histogram in Fig.~\ref{fig:norm_wait_times}, is a smooth function that clearly deviates from an exponential.

Once we have normalized the waiting times for each star across yearlong bins, we then combine all stars within a given stellar sample as normalized wait times.

The WTD is separated into short wait times, which are a combination of sympathetic flaring and Poisson randomness, and longer wait times, which are a priori solely arising from a Poisson process. To define short wait times, we systematically histogram the WTD with 100 different bin widths. The aim was to determine the optimum bin width such that all potentially sympathetic flares are isolated to the first bin. We take the optimal bin width to be the one where the second histogram bin best matched an exponential distribution fit to the entire WTD except for the first bin. When we resample our flare catalogs, we similarly recalculate the bin widths for each resampling.

In calculating the percent of excess flares, we use two methods, with the two different models of WTDs mentioned above. In the first method, we calculate the excess at short wait times by calculating the area contained by the lowest bin of each bootstrapped WTD, and then we subtract out the area from the assumed Poisson WTD, similarly calculated with the \texttt{quad} function from the \texttt{scipy.integrate} function. To extend to longer wait times, we take that calculated difference in the lowest bin and divide by the total area under the histogram. 

In the second method, we fit a Weibull distribution to the data. We take the $\lambda$ value from the exponential which is best-fit from the Poisson data (i.e. the WTD excluding the first bin) and we let $k$ vary to find the best fit. We employ the \texttt{quad} function from the \texttt{scipy.integrate} function to determine the area underneath the best-fit Weibull distribution, and then we subtract the area from the assumed exponential WTD. For short wait times, we limit our integral from the minimum normalized waiting time to the edge of the first bin, as defined in the paragraph above. For the total percent excess, our integral bounds extended from the minimum to maximum normalized waiting time for each sample. 

For both methods of excess determination, we repeat the process 1000 times on bootstrapped WTDs.
\textbf{}

\section{Results}\label{sec: results}

\subsection{The Discovery of Sympathetic Flaring}\label{subsec:sympathetic_discovery}

In Fig.~\ref{fig: WTDs} we plot the normalized flare wait times for all four flare catalogs, the three compiled by \textsc{toffee}, and the one compiled by \texttt{stella}. There is strong evidence for sympathetic flaring: the majority of the WTD is well fit by an exponential, except at short wait times where there is a significant excess compared with a Poisson process. We separate the WTD into two parts: short wait times (sympathetic \& Poisson) that are constrained to the lowest bin and longer wait times (Poisson only). Then, we fit an exponential to Poisson-only data and extrapolated it to shorter wait times, as discussed in Sect.~\ref{sec:stats}. 

To quantify the rate of sympathetic flaring, we calculated the fraction of flares in the shortest wait time bin that fall above the fitted exponential distribution. We repeat this process on the 1000 bootstrapped samples of flare wait times. This method, denoted as ``Poisson excess,'' reveals that between $\sim$25 and 65\% of short wait time flares are sympathetic (Fig.~\ref{fig: excess} top left) and $\sim$4-8\% of all flares are sympathetic (Fig.~\ref{fig: excess}, top right).

When fitting the Weibull distribution, we fix $\lambda$ rate parameter in the same way as with the Poisson distribution, as the value of the fitted exponential to the Poisson-only WTD, and then we find the best-fitting $k$ across the entire WTD. In Fig.~\ref{fig: WTDs} we show the best-fitting Weibull distribution (solid gray line). From bootstrapping, we calculate a standard deviation for $k$ in all four samples: Seli (\textsc{toffee}) $k= 0.910\pm 0.006$; Yudovich (\textsc{toffee}) $k=0.902\pm0.009$; Feinstein (\textsc{toffee}) $k=0.894\pm0.009$; and Feinstein (\texttt{stella}) $k=0.878\pm0.012$. Since in each sample $k$ is statistically significantly less than 1, our results demonstrate an increased abundance of short wait time events compared with an exponential distribution, as expected for sympathetic flaring. Visually, the Weibull is indistinguishable from an exponential (dashed gray line) at long wait times, but the Weibull hooks upwards at short wait times.

We also use the Weibull distribution as an alternative means of quantifying sympathetic flaring, calculating the percentage of short wait time flares that are sympathetic by integrating the Weibull function over the first histogram bin and subtracting the integral of the exponential function over the same bin (Fig.~\ref{fig: excess}, bottom left). According to this method, $\sim$ 18-30\% of short wait-time flares are sympathetic.  To get the percentage of all flares that are sympathetic, we repeat this process but over the entire WTD (Fig.~\ref{fig: excess}, bottom right). Similarly to our first method, $\sim$ 4-8\% of all flares are sympathetic, matching the solar value.

Even though the Weibull distribution is a significantly better fit than an exponential, it is visually not a perfect match with the observations. Across all four flare catalogs, the Weibull consistently underestimates the amount of shortest wait-times but slightly overestimates the longer wait-times. This feature indicates that our potential estimate that $\sim$18-30\% of short wait-time flares are sympathetic is likely an underestimate. 

In comparing the two methods of quantifying sympathetic excess, there is a difference between the percentages at short wait times specifically. This inconsistency is a result of the disparate behavior of the Weibull PDF and the exponential PDF as well as the difference in calculation. For all of our results, as mentioned above, the Weibull underestimates the amount of shortest wait-times. In contrast, when calculating the Poisson excess percentage, we are able to use the full area of the shortest wait-time bin. Hence, the Poisson excess percentage at short wait times is much higher than the Weibull excess. The percentage of sympathetic flares across all flares, moreover, is roughly the same between the two methods because the Weibull distribution slightly overestimates the longer wait-time probabilities and thus undoes the underestimation from earlier.

\subsection{Comparison to the Sun}

\begin{figure}
\centering
\includegraphics[width=\linewidth]
{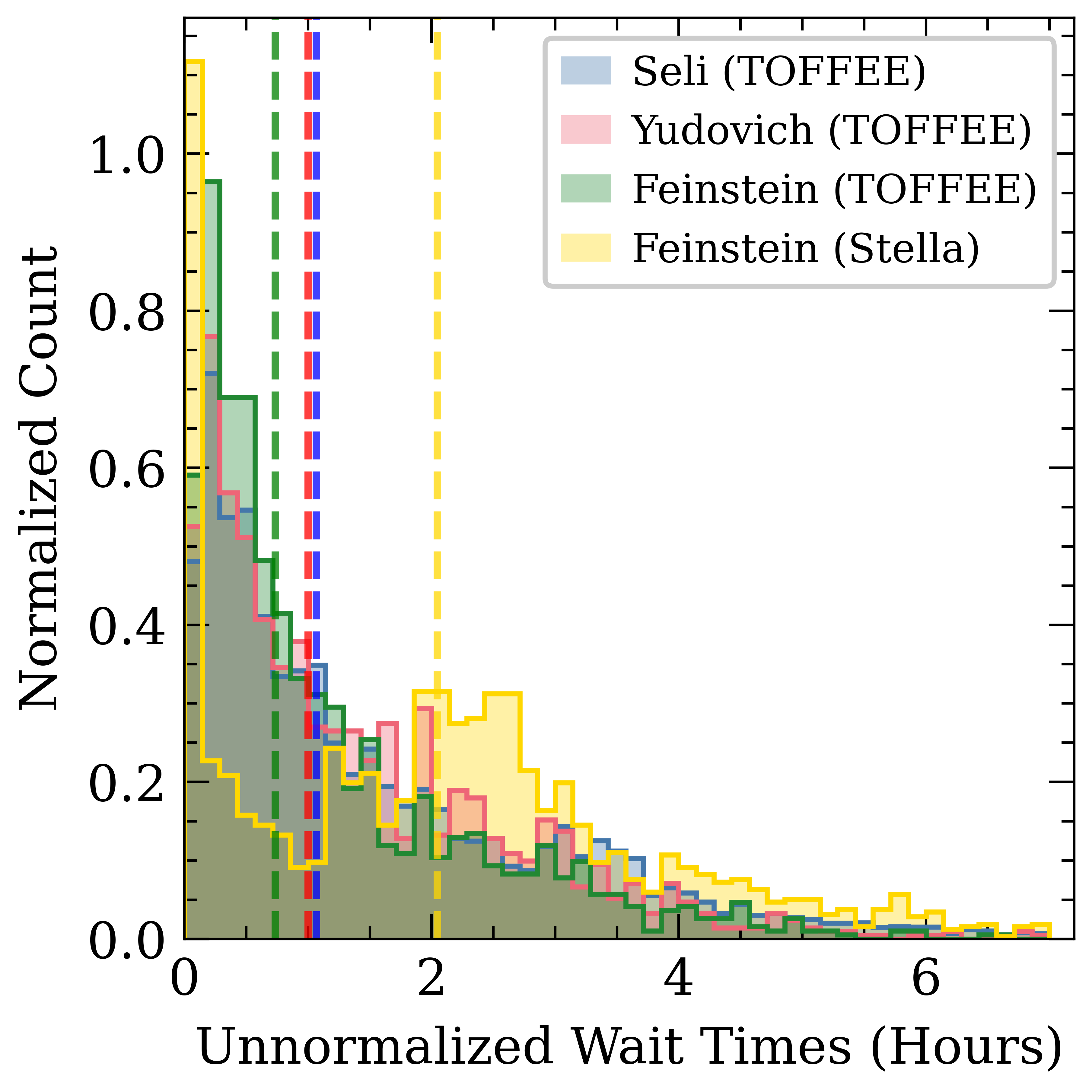}
\caption{The distribution of the unnormalized wait times of all flares deemed potentially sympathetic from each of our samples. The medians of the respective distributions are shown with the dashed lines that match the colors in the histogram: Seli (blue), Yudovich (pink), Feinstein by \textsc{toffee} (green) and Feinstein by \texttt{stella} (yellow).}\label{fig: sympathetic wait times}
\end{figure}

In Fig.~\ref{fig: sympathetic wait times} we plot the absolute wait time, in hours, of all flares that fall into the first histogram bin and are hence considered potentially sympathetic. We see the medians range from between 0.5 and 2 hours -- with the samples generated from \textsc{toffee} having medians between 0.5 and 1.5 hours -- similar to the Sun \citep{2022Mawad, 2025Guite}. The wait times for each \textsc{toffee} sample peak at the lowest wait times as well, highlighting \textsc{toffee}'s completeness in detecting overlapping flares.  

Ultimately, our results demonstrate that sympathetic flares on other stars occur at similar rates, between 4\% and 8\% overall, and at similar times as the Sun. 

\subsection{Comparison to Other Flare Samples}\label{subsec:comparison_other_samples}

\citet{2024Feinstein}, \citet{Seli2025}, and \citet{2025Yudovich} provide both a sample of stars and a sample of flares. Compared to the original 26,355 flares from the original \cite{2020Feinstein}, we recovered less flares due to the \texttt{stella} code being capable to find numerous low amplitude flares less than $3\sigma$ above the median flux. The original flare sample from \cite{2025Yudovich} found 10,142 flares, with 4,000 more flares found by \textsc{toffee}. The original \cite{Seli2025} sample contained 121,895, with \textsc{toffee} finding significantly more. However, whilst we use the three stellar samples, the only flare sample we use in our analysis is from \citet{2024Feinstein}, as discovered by \textsc{stella}. This choice was made because the threshold detection methods implemented in \citet{Seli2025} and \citet{2025Yudovich} are insensitive to closely-separated, overlapping flares. 

To demonstrate this, we perform a Weibull regression on all three samples, and we find a value of $k = 0.878 \pm 0.012$, $k = 1.12 \pm 0.01$, $k = 1.21 \pm 0.004$ for the given flare catalogs from Feinstein, Yudovich, and Seli, respectively. Recall that $k>1$ indicates a \textit{dearth} of short wait-times relative to an exponential WTD. Otherwise put, the Yudovich and Seli samples find substantially fewer flares than expected from stochastic flare occurrence and thus are incomplete \citep{Seli2025, 2025Yudovich}.

We also test what would happen if we used \textsc{toffee} but without secondary flare detection. After removing all classified secondary flares, we find Weibull functions with a best fit value of $k = 1.06 \pm 0.01$, $k = 1.02 \pm 0.01$, $k = 1.041 \pm 0.005$ for Feinstein, Yudovich, and Seli, respectively. This demonstrates that the ability to detect secondary flares is essential for probing sympathetic flares, but also that \textsc{toffee} is highly complete across primary flares.

\section{Summary and Conclusions}\label{sec:conc}

In this study we discovered over 200,000 flares on over 16,000 stars, using a custom and publicly available flare detection algorithm \textsc{toffee} that is sensitive to flares that are closely separated in time. By carefully analyzing the wait time distribution, we discover an excess of temporarily close separation flares, which we take as a strong signature of sympathetic flaring. This is the first time sympathetic flaring has been seen on other stars with statistical robustness. The rate of sympathetic flares is between $\sim 4-9\%$, which is similar to the 5\% rate observed on the Sun. The sympathetic flares in our sample are typically separated by $\sim 0.5-1.5$ hours, which is also similar to the rate seen on the Sun.

Furthermore, the majority of our flares come from M-dwarfs, and hence sympathetic flaring is something that occurs across vastly different spectral types. Although the underlying physical mechanism  of sympathetic flaring is unknown, we now know that the phenomenon is not limited to the Sun.

The \textsc{toffee} code is currently available through pip installation with documentation on the \textsc{toffee} GitHub page \footnote{\href{https://github.com/JasonReeves702/TOFFEE}{https://github.com/JasonReeves702/TOFFEE}} with the three flare catalogs utilized in this study found within the repository \footnote{\href{https://github.com/JasonReeves702/TOFFEE/tree/main/Flare_Catalogs}{Flare\_Catalogs}}.

 \section*{Acknowledgments}

The authors thank an anonymous referee for comments which significantly improved this paper. This work began as a class project in AST-51/151 ``Astrophysics Laboratory'' in Spring 2025 at Tufts University. We appreciate the hard work of all of the students in the class. A.Z. was sponsored by the Tufts Summer Scholars program during Summer 2025.

\bibliography{science_template} 

@ARTICLE{Moreton1960,
       author = {{Moreton}, G.~E.},
        title = "{H{\ensuremath{\alpha}} Observations of Flare-Initiated Disturbances with Velocities \raisebox{-0.5ex}\textasciitilde1000 km/sec.}",
      journal = {\aj},
         year = 1960,
        month = jan,
       volume = {65},
        pages = {494},
          doi = {10.1086/108346},
       adsurl = {https://ui.adsabs.harvard.edu/abs/1960AJ.....65U.494M}
}

@ARTICLE{Uchida1968,
       author = {{Uchida}, Yutaka},
        title = "{Propagation of Hydromagnetic Disturbances in the Solar Corona and Moreton's Wave Phenomenon}",
      journal = {\solphys},
         year = 1968,
        month = may,
       volume = {4},
       number = {1},
        pages = {30-44},
          doi = {10.1007/BF00146996},
       adsurl = {https://ui.adsabs.harvard.edu/abs/1968SoPh....4...30U}
}

@article{2018Doyle,
    author = {Doyle, L and Ramsay, G and Doyle, J G and Wu, K and Scullion, E},
    title = {Investigating the rotational phase of stellar flares on M dwarfs using K2 short cadence data},
    journal = {Monthly Notices of the Royal Astronomical Society},
    volume = {480},
    number = {2},
    pages = {2153-2164},
    year = {2018},
    month = {07},
    issn = {0035-8711},
    doi = {10.1093/mnras/sty1963},
    url = {https://doi.org/10.1093/mnras/sty1963},
    eprint = {https://academic.oup.com/mnras/article-pdf/480/2/2153/25442643/sty1963.pdf},
}

@article{2019Doyle,
    author = {Doyle, L and Ramsay, G and Doyle, J G and Wu, K},
    title = {Probing the origin of stellar flares on M dwarfs using TESS data sectors 1–3},
    journal = {Monthly Notices of the Royal Astronomical Society},
    volume = {489},
    number = {1},
    pages = {437-445},
    year = {2019},
    month = {08},
    issn = {0035-8711},
    doi = {10.1093/mnras/stz2205},
    url = {https://doi.org/10.1093/mnras/stz2205},
    eprint = {https://academic.oup.com/mnras/article-pdf/489/1/437/29201307/stz2205.pdf},
}

@article{2020Doyle,
    author = {Doyle, L and Ramsay, G and Doyle, J G},
    title = {Superflares and variability in solar-type stars with TESS in the Southern hemisphere},
    journal = {Monthly Notices of the Royal Astronomical Society},
    volume = {494},
    number = {3},
    pages = {3596-3610},
    year = {2020},
    month = {04},
    issn = {0035-8711},
    doi = {10.1093/mnras/staa923},
    url = {https://doi.org/10.1093/mnras/staa923},
    eprint = {https://academic.oup.com/mnras/article-pdf/494/3/3596/33145067/staa923.pdf},
}

@article{2015Lurie,
   title={KEPLERFLARES III: STELLAR ACTIVITY ON GJ 1245A AND B},
   volume={800},
   ISSN={1538-4357},
   url={http://dx.doi.org/10.1088/0004-637X/800/2/95},
   DOI={10.1088/0004-637x/800/2/95},
   number={2},
   journal={The Astrophysical Journal},
   publisher={American Astronomical Society},
   author={Lurie, John C. and Davenport, James R. A. and Hawley, Suzanne L. and Wilkinson, Tessa D. and Wisniewski, John P. and Kowalski, Adam F. and Hebb, Leslie},
   year={2015},
   month=feb, pages={95} }

@article{2014Hawley,
   title={KEPLERFLARES. I. ACTIVE AND INACTIVE M DWARFS},
   volume={797},
   ISSN={1538-4357},
   url={http://dx.doi.org/10.1088/0004-637X/797/2/121},
   DOI={10.1088/0004-637x/797/2/121},
   number={2},
   journal={The Astrophysical Journal},
   publisher={American Astronomical Society},
   author={Hawley, Suzanne L. and Davenport, James R. A. and Kowalski, Adam F. and Wisniewski, John P. and Hebb, Leslie and Deitrick, Russell and Hilton, Eric J.},
   year={2014},
   month=dec, pages={121} }

@ARTICLE{HuntWalker2012,
       author = {{Hunt-Walker}, Nicholas M. and {Hilton}, Eric J. and {Kowalski}, Adam F. and {Hawley}, Suzanne L. and {Matthews}, Jaymie M.},
        title = "{MOST Observations of the Flare Star AD Leo}",
      journal = {\pasp},
         year = 2012,
        month = jun,
       volume = {124},
       number = {916},
        pages = {545},
          doi = {10.1086/666495},
archivePrefix = {arXiv},
       eprint = {1206.5019},
 primaryClass = {astro-ph.SR},
       adsurl = {https://ui.adsabs.harvard.edu/abs/2012PASP..124..545H}
}

@ARTICLE{Warmuth2004,
       author = {{Warmuth}, A. and {Vr{\v{s}}nak}, B. and {Magdaleni{\'c}}, J. and {Hanslmeier}, A. and {Otruba}, W.},
        title = "{A multiwavelength study of solar flare waves. II. Perturbation characteristics and physical interpretation}",
      journal = {\aap},
         year = 2004,
        month = may,
       volume = {418},
        pages = {1117-1129},
          doi = {10.1051/0004-6361:20034333},
       adsurl = {https://ui.adsabs.harvard.edu/abs/2004A&A...418.1117W}
}

@misc{Paegert2022,
       author = {{Paegert}, M. and {Stassun}, K.~G. and {Collins}, K.~A. and {Pepper}, J. and {Torres}, G. and {Jenkins}, J. and {Twicken}, J.~D. and {Latham}, D.~W.},
        title = "{VizieR Online Data Catalog: TESS Input Catalog version 8.2 (TIC v8.2) (Paegert+, 2021)}",
 howpublished = {VizieR On-line Data Catalog: IV/39.  Originally published in: 2021arXiv210804778P},
         year = 2022,
        month = feb,
          eid = {IV/39},
       adsurl = {https://ui.adsabs.harvard.edu/abs/2022yCat.4039....0P}
}

@INPROCEEDINGS{Wheatland2000,
       author = {{Wheatland}, M.~S.},
        title = "{The Origin of the Solar Flare Waiting-time Distribution}",
    booktitle = {AAS/Solar Physics Division Meeting \#31},
         year = 2000,
       series = {AAS/Solar Physics Division Meeting},
       volume = {31},
        month = oct,
          eid = {02.56},
        pages = {02.56},
       adsurl = {https://ui.adsabs.harvard.edu/abs/2000SPD....31.0256W}
}

@INPROCEEDINGS{2024Davenport,
       author = {{Davenport}, James and {Tovar Mendoza}, Guadalupe and {Wainer}, Tobin},
        title = "{Searching for Activity Cycles using Stellar Flares}",
    booktitle = {American Astronomical Society Meeting Abstracts},
         year = 2024,
       series = {American Astronomical Society Meeting Abstracts},
       volume = {243},
        month = feb,
          eid = {355.01},
        pages = {355.01},
       adsurl = {https://ui.adsabs.harvard.edu/abs/2024AAS...24335501D}
}

@ARTICLE{2024Paudel,
       author = {{Paudel}, Rishi R. and {Barclay}, Thomas and {Youngblood}, Allison and {Quintana}, Elisa V. and {Schlieder}, Joshua E. and {Vega}, Laura D. and {Gilbert}, Emily A. and {Osten}, Rachel A. and {Peacock}, Sarah and {Tristan}, Isaiah I. and {Feliz}, Dax L. and {Boyd}, Patricia T. and {Davenport}, James R.~A. and {Huber}, Daniel and {Kowalski}, Adam F. and {Monsue}, Teresa and {Silverstein}, Michele L.},
        title = "{A Multiwavelength Survey of Nearby M Dwarfs: Optical and Near-ultraviolet Flares and Activity with Contemporaneous TESS, Kepler/K2, Swift, and HST Observations}",
      journal = {\apj},
         year = 2024,
        month = aug,
       volume = {971},
       number = {1},
          eid = {24},
        pages = {24},
          doi = {10.3847/1538-4357/ad487d},
archivePrefix = {arXiv},
       eprint = {2404.12310},
 primaryClass = {astro-ph.SR},
       adsurl = {https://ui.adsabs.harvard.edu/abs/2024ApJ...971...24P}
}

@ARTICLE{2024Wainer,
       author = {{Wainer}, Tobin M. and {Davenport}, James R.~A. and {Tovar Mendoza}, Guadalupe and {Feinstein}, Adina D. and {Wagg}, Tom},
        title = "{Searching for Stellar Activity Cycles Using Flares: The Short- and Long-timescale Activity Variations of TIC-272272592}",
      journal = {\aj},
         year = 2024,
        month = dec,
       volume = {168},
       number = {6},
          eid = {232},
        pages = {232},
          doi = {10.3847/1538-3881/ad7bb2},
archivePrefix = {arXiv},
       eprint = {2409.06631},
 primaryClass = {astro-ph.SR},
       adsurl = {https://ui.adsabs.harvard.edu/abs/2024AJ....168..232W}
}

@ARTICLE{2022Crowley,
       author = {{Crowley}, James and {Wheatland}, Michael S. and {Yang}, Kai},
        title = "{Observed Rate Variations in Superflaring G-type Stars}",
      journal = {\apj},
         year = 2022,
        month = dec,
       volume = {941},
       number = {2},
          eid = {193},
        pages = {193},
          doi = {10.3847/1538-4357/aca476},
archivePrefix = {arXiv},
       eprint = {2212.00993},
 primaryClass = {astro-ph.SR},
       adsurl = {https://ui.adsabs.harvard.edu/abs/2022ApJ...941..193C}
}

@ARTICLE{2002WheatlandLitvinenko,
       author = {{Wheatland}, M.~S. and {Litvinenko}, Y.~E.},
        title = "{Understanding Solar Flare Waiting-Time Distributions}",
      journal = {\solphys},
         year = 2002,
        month = dec,
       volume = {211},
       number = {1},
        pages = {255-274},
          doi = {10.1023/A:1022430308641},
       adsurl = {https://ui.adsabs.harvard.edu/abs/2002SoPh..211..255W}
}

@ARTICLE{2014Tellonietal,
       author = {{Telloni}, D. and {Carbone}, V. and {Lepreti}, F. and {Antonucci}, E.},
        title = "{Stochasticity and Persistence of Solar Coronal Mass Ejections}",
      journal = {\apjl},
         year = 2014,
        month = jan,
       volume = {781},
       number = {1},
          eid = {L1},
        pages = {L1},
          doi = {10.1088/2041-8205/781/1/L1},
       adsurl = {https://ui.adsabs.harvard.edu/abs/2014ApJ...781L...1T}
}

@ARTICLE{2022Mawad,
       author = {{Mawad}, Ramy and {Moussas}, Xenophon},
        title = "{Sympathetic solar flare: characteristics and homogeneities}",
      journal = {\apss},
         year = 2022,
        month = nov,
       volume = {367},
       number = {11},
          eid = {107},
        pages = {107},
          doi = {10.1007/s10509-022-04145-3},
       adsurl = {https://ui.adsabs.harvard.edu/abs/2022Ap&SS.367..107M}
}

@ARTICLE{2025Guite,
       author = {{Guit{\'e}}, L. -S. and {Strugarek}, A. and {Charbonneau}, P.},
        title = "{Flaring together: A preferred angular separation between sympathetic flares on the Sun}",
      journal = {\aap},
         year = 2025,
        month = feb,
       volume = {694},
          eid = {A74},
        pages = {A74},
          doi = {10.1051/0004-6361/202452381},
archivePrefix = {arXiv},
       eprint = {2412.10143},
 primaryClass = {astro-ph.SR},
       adsurl = {https://ui.adsabs.harvard.edu/abs/2025A&A...694A..74G}
}

@ARTICLE{Moon2002,
       author = {{Moon}, Y. -J. and {Choe}, G.~S. and {Park}, Y.~D. and {Wang}, Haimin and {Gallagher}, Peter T. and {Chae}, Jongchul and {Yun}, H.~S. and {Goode}, Philip R.},
        title = "{Statistical Evidence for Sympathetic Flares}",
      journal = {\apj},
         year = 2002,
        month = jul,
       volume = {574},
       number = {1},
        pages = {434-439},
          doi = {10.1086/340945},
       adsurl = {https://ui.adsabs.harvard.edu/abs/2002ApJ...574..434M}
}

@ARTICLE{Hakamata2025,
       author = {{Hakamata}, Tomohiro and {Matsumoto}, Hironori and {Odaka}, Hirokazu and {Takasao}, Shinsuke},
        title = "{NuSTAR detection of a hot stellar superflare with a temperature of 95 MK in hard X-rays}",
      journal = {\pasj},
         year = 2025,
        month = apr,
       volume = {77},
       number = {2},
        pages = {356-369},
          doi = {10.1093/pasj/psaf002},
archivePrefix = {arXiv},
       eprint = {2501.16710},
 primaryClass = {astro-ph.SR},
       adsurl = {https://ui.adsabs.harvard.edu/abs/2025PASJ...77..356H}
}

@ARTICLE{Li2023,
       author = {{Li}, Guang-Wei and {Wu}, Chao and {Zhou}, Gui-Ping and {Yang}, Chen and {Li}, Hua-Li and {Chen}, Jie and {Xin}, Li-Ping and {Wang}, Jing and {Haerken}, Hasitieer and {Ma}, Chao-Hong and {Cai}, Hong-Bo and {Han}, Xu-Hui and {Huang}, Lei and {Lu}, Xiao-Meng and {Bai}, Jian-Ying and {Zhang}, Xu-Kang and {Hao}, Xin-Li and {Wang}, Xiang-Yu and {Dai}, Zi-Gao and {Liang}, En-Wei and {Meng}, Xiao-Feng and {Wei}, Jian-Yan},
        title = "{Magnetic Activity and Parameters of 43 Flare Stars in the GWAC Archive}",
      journal = {Research in Astronomy and Astrophysics},
         year = 2023,
        month = jan,
       volume = {23},
       number = {1},
          eid = {015016},
        pages = {015016},
          doi = {10.1088/1674-4527/aca506},
archivePrefix = {arXiv},
       eprint = {2211.11240},
 primaryClass = {astro-ph.SR},
       adsurl = {https://ui.adsabs.harvard.edu/abs/2023RAA....23a5016L}
}

@ARTICLE{Zhang2025,
       author = {{Zhang}, Andy B. and {Reeves}, Jason R. and {Martin}, David V. and {Pratt}, Veronica and {Tubthong}, Wata and {Weinstein}, Arielle and {Ward}, Isabella E.},
        title = "{Starspots and Flares are Generally Not Correlated}",
      journal = {arXiv e-prints},
         year = 2025,
        month = nov,
          eid = {arXiv:2512.01051},
        pages = {arXiv:2512.01051},
          doi = {10.48550/arXiv.2512.01051},
archivePrefix = {arXiv},
       eprint = {2512.01051},
 primaryClass = {astro-ph.SR},
       adsurl = {https://ui.adsabs.harvard.edu/abs/2025arXiv251201051Z}
}

@ARTICLE{2018Li,
       author = {{Li}, C. and {Zhong}, S.~J. and {Xu}, Z.~G. and {He}, H. and {Yan}, Y. and {Chen}, P.~F. and {Fang}, C.},
        title = "{Waiting time distributions of solar and stellar flares: Poisson process or with memory?}",
      journal = {\mnras},
         year = 2018,
        month = sep,
       volume = {479},
       number = {1},
        pages = {L139-L142},
          doi = {10.1093/mnrasl/sly117},
       adsurl = {https://ui.adsabs.harvard.edu/abs/2018MNRAS.479L.139L}
}

@ARTICLE{2024Feinstein,
       author = {{Feinstein}, Adina D. and {Seligman}, Darryl Z. and {France}, Kevin and {Gagn{\'e}}, Jonathan and {Kowalski}, Adam},
        title = "{Evolution of Flare Activity in GKM Stars Younger Than 300 Myr over Five Years of TESS Observations}",
      journal = {\aj},
         year = 2024,
        month = aug,
       volume = {168},
       number = {2},
          eid = {60},
        pages = {60},
          doi = {10.3847/1538-3881/ad4edf},
archivePrefix = {arXiv},
       eprint = {2405.00850},
 primaryClass = {astro-ph.SR},
       adsurl = {https://ui.adsabs.harvard.edu/abs/2024AJ....168...60F}
}

@ARTICLE{2025Yudovich,
       author = {{Yudovich}, Denise G. and {Yang}, Kai E. and {Sun}, Xudong},
        title = "{Analyzing the Morphology of Late-phase Stellar Flares from G-, K-, and M-type Stars}",
      journal = {\apj},
         year = 2025,
        month = may,
       volume = {984},
       number = {2},
          eid = {186},
        pages = {186},
          doi = {10.3847/1538-4357/adc695},
archivePrefix = {arXiv},
       eprint = {2503.16181},
 primaryClass = {astro-ph.SR},
       adsurl = {https://ui.adsabs.harvard.edu/abs/2025ApJ...984..186Y}
}

@ARTICLE{Seli2025,
       author = {{Seli}, B. and {Vida}, K. and {Ol{\'a}h}, K. and {G{\"o}rgei}, A. and {So{\'o}s}, Sz. and {P{\'a}l}, A. and {Kriskovics}, L. and {K{\H{o}}v{\'a}ri}, Zs.},
        title = "{Stellar flare morphology with TESS across the main sequence}",
      journal = {\aap},
         year = 2025,
        month = feb,
       volume = {694},
          eid = {A161},
        pages = {A161},
          doi = {10.1051/0004-6361/202452489},
archivePrefix = {arXiv},
       eprint = {2412.12989},
 primaryClass = {astro-ph.SR},
       adsurl = {https://ui.adsabs.harvard.edu/abs/2025A&A...694A.161S}
}

@article{2014Davenport,
doi = {10.1088/0004-637X/797/2/122},
url = {https://dx.doi.org/10.1088/0004-637X/797/2/122},
year = {2014},
month = {dec},
publisher = {The American Astronomical Society},
volume = {797},
number = {2},
pages = {122},
author = {Davenport, James R. A. and Hawley, Suzanne L. and Hebb, Leslie and Wisniewski, John P. and Kowalski, Adam F. and Johnson, Emily C. and Malatesta, Michael and Peraza, Jesus and Keil, Marcus and Silverberg, Steven M. and Jansen, Tiffany C. and Scheffler, Matthew S. and Berdis, Jodi R. and Larsen, Daniel M. and Hilton, Eric J.},
title = {KEPLER FLARES. II. THE TEMPORAL MORPHOLOGY OF WHITE-LIGHT FLARES ON GJ 1243},
journal = {The Astrophysical Journal}
}

@ARTICLE{FritzovaSvestkova1976,
       author = {{Fritzova-Svestkova}, L. and {Chase}, R.~C. and {Svestka}, Z.},
        title = "{On the occurrence of sympathetic flares.}",
      journal = {\solphys},
         year = 1976,
        month = jun,
       volume = {48},
       number = {2},
        pages = {275-286},
          doi = {10.1007/BF00151996},
       adsurl = {https://ui.adsabs.harvard.edu/abs/1976SoPh...48..275F}
}

@ARTICLE{2021Ilin,
       author = {{Ilin}, Ekaterina and {Schmidt}, Sarah J. and {Poppenh{\"a}ger}, Katja and {Davenport}, James R.~A. and {Kristiansen}, Martti H. and {Omohundro}, Mark},
        title = "{Flares in open clusters with K2. II. Pleiades, Hyades, Praesepe, Ruprecht 147, and M 67}",
      journal = {\aap},
         year = 2021,
        month = jan,
       volume = {645},
          eid = {A42},
        pages = {A42},
          doi = {10.1051/0004-6361/202039198},
archivePrefix = {arXiv},
       eprint = {2010.05576},
 primaryClass = {astro-ph.SR},
       adsurl = {https://ui.adsabs.harvard.edu/abs/2021A&A...645A..42I}
}

@ARTICLE{2020Feinstein,
       author = {{Feinstein}, Adina D. and {Montet}, Benjamin T. and {Ansdell}, Megan and {Nord}, Brian and {Bean}, Jacob L. and {G{\"u}nther}, Maximilian N. and {Gully-Santiago}, Michael A. and {Schlieder}, Joshua E.},
        title = "{Flare Statistics for Young Stars from a Convolutional Neural Network Analysis of TESS Data}",
      journal = {\aj},
         year = 2020,
        month = nov,
       volume = {160},
       number = {5},
          eid = {219},
        pages = {219},
          doi = {10.3847/1538-3881/abac0a},
archivePrefix = {arXiv},
       eprint = {2005.07710},
 primaryClass = {astro-ph.SR},
       adsurl = {https://ui.adsabs.harvard.edu/abs/2020AJ....160..219F}
}

@article{2014Pitkin,
    author = {Pitkin, M. and Williams, D. and Fletcher, L. and Grant, S. D. T.},
    title = {A Bayesian method for detecting stellar flares},
    journal = {Monthly Notices of the Royal Astronomical Society},
    volume = {445},
    number = {3},
    pages = {2268-2284},
    year = {2014},
    month = {10},
    issn = {0035-8711},
    doi = {10.1093/mnras/stu1889},
    url = {https://doi.org/10.1093/mnras/stu1889},
    eprint = {https://academic.oup.com/mnras/article-pdf/445/3/2268/13766423/stu1889.pdf},
}

@article{2022Howard,
doi = {10.3847/1538-4357/ac426e},
url = {https://dx.doi.org/10.3847/1538-4357/ac426e},
year = {2022},
month = {feb},
publisher = {The American Astronomical Society},
volume = {926},
number = {2},
pages = {204},
author = {Howard, Ward S. and MacGregor, Meredith A.},
title = {No Such Thing as a Simple Flare: Substructure and Quasi-periodic Pulsations Observed in a Statistical Sample of 20 s Cadence TESS Flares},
journal = {The Astrophysical Journal}
}

@ARTICLE{2023Rivera,
       author = {{Rivera}, Elmer C. and {Johnson}, Jay R. and {Homan}, Jonathan and {Wing}, Simon},
        title = "{How noise thresholds affect the information content of stellar flare sequences}",
      journal = {\aap},
         year = 2023,
        month = feb,
       volume = {670},
          eid = {A143},
        pages = {A143},
          doi = {10.1051/0004-6361/202245309},
       adsurl = {https://ui.adsabs.harvard.edu/abs/2023A&A...670A.143R}
}

@ARTICLE{2015Chang,
       author = {{Chang}, S. -W. and {Byun}, Y. -I. and {Hartman}, J.~D.},
        title = "{Photometric Study on Stellar Magnetic Activity. I. Flare Variability of Red Dwarf Stars in the Open Cluster M37}",
      journal = {\apj},
         year = 2015,
        month = nov,
       volume = {814},
       number = {1},
          eid = {35},
        pages = {35},
          doi = {10.1088/0004-637X/814/1/35},
archivePrefix = {arXiv},
       eprint = {1510.01005},
 primaryClass = {astro-ph.SR},
       adsurl = {https://ui.adsabs.harvard.edu/abs/2015ApJ...814...35C}
}

@ARTICLE{2019Yang,
       author = {{Yang}, Huiqin and {Liu}, Jifeng},
        title = "{The Flare Catalog and the Flare Activity in the Kepler Mission}",
      journal = {\apjs},
         year = 2019,
        month = apr,
       volume = {241},
       number = {2},
          eid = {29},
        pages = {29},
          doi = {10.3847/1538-4365/ab0d28},
archivePrefix = {arXiv},
       eprint = {1903.01056},
 primaryClass = {astro-ph.SR},
       adsurl = {https://ui.adsabs.harvard.edu/abs/2019ApJS..241...29Y}
}

@ARTICLE{2022Ward,
       author = {{Howard}, Ward S. and {MacGregor}, Meredith A.},
        title = "{No Such Thing as a Simple Flare: Substructure and Quasi-periodic Pulsations Observed in a Statistical Sample of 20 s Cadence TESS Flares}",
      journal = {\apj},
         year = 2022,
        month = feb,
       volume = {926},
       number = {2},
          eid = {204},
        pages = {204},
          doi = {10.3847/1538-4357/ac426e},
archivePrefix = {arXiv},
       eprint = {2110.13155},
 primaryClass = {astro-ph.SR},
       adsurl = {https://ui.adsabs.harvard.edu/abs/2022ApJ...926..204H}
}

@book{1977Tukey,
  title={Data Analysis and Regression: A Second Course in Statistics},
  author={Mosteller, F. and Tukey, J.W.},
  isbn={9780201048544},
  lccn={lc76015465},
  series={Addison-Wesley series in behavioral science},
  url={https://books.google.com/books?id=n4dYAAAAMAAJ},
  year={1977},
  publisher={Addison-Wesley Publishing Company}
}

@ARTICLE{2019Hippke,
       author = {{Hippke}, Michael and {David}, Trevor J. and {Mulders}, Gijs D. and {Heller}, Ren{\'e}},
        title = "{W{\={o}}tan: Comprehensive Time-series Detrending in Python}",
      journal = {\aj},
         year = 2019,
        month = oct,
       volume = {158},
       number = {4},
          eid = {143},
        pages = {143},
          doi = {10.3847/1538-3881/ab3984},
archivePrefix = {arXiv},
       eprint = {1906.00966},
 primaryClass = {astro-ph.EP},
       adsurl = {https://ui.adsabs.harvard.edu/abs/2019AJ....158..143H}
}
\bibliographystyle{sciencemag}

\clearpage 





\end{document}